\documentclass[3p, review, authoryear, 10pt]{elsarticle}

\usepackage{xcolor}
\usepackage[english]{babel}
\usepackage{eurosym}
\usepackage{setspace}
\usepackage{bm}
\usepackage{amsfonts}
\usepackage{natbib}
 \bibpunct[, ]{(}{)}{,}{a}{}{,}%

\usepackage{threeparttable}
\usepackage{graphicx}
\usepackage{verbatim}
\usepackage{float}
\restylefloat{table}
\usepackage{mathtools}
\usepackage{wrapfig}
\usepackage{mdwlist}
\usepackage{listings}
\usepackage{url}
\usepackage{color}
\usepackage{xcolor}
\usepackage{booktabs}
\usepackage{caption}

\usepackage{chngcntr}
\usepackage{subfig}
\usepackage{adjustbox}
\usepackage{enumerate}
\usepackage{hyperref}
\usepackage{enumitem}
\usepackage{float}
\usepackage{microtype}
\usepackage{ulem}
\usepackage{pdfpages}
\usepackage{epstopdf}
\usepackage{longtable, tabu}

\begin{document}

\begin{frontmatter}

\journal{Renewable and Sustainable Energy Reviews}

\title{Green Hydrogen Plant: Optimal control strategies for integrated  hydrogen storage and power generation with wind energy}

\author[groningen]{Arjen T.~Veenstra} \ead{a.t.veenstra@rug.nl} 
\author[eindhoven]{Albert H.~Schrotenboer} \ead{a.h.schrotenboer@tue.nl} 
\author[groningen]{Michiel A.~J.~uit het Broek} \ead{a.j.uit.het.broek@rug.nl}
\author[groningen]{Evrim Ursavas} \ead{e.ursavas@rug.nl} 

\cortext[corrauthor]{Corresponding author: Nettelbosje 2, 9747 AE Groningen, the Netherlands.}

\address[groningen]{Department of Operations, University of Groningen, PO Box 800, 9700 AV Groningen, the Netherlands}
\address[eindhoven]{Department of Operations, Planning, Accounting, and Control, School of Industrial Engineering, 5600 MB, Eindhoven University of Technology}

\begin{abstract}
\small

The intermittent nature of renewable energy resources such as wind and solar causes the energy supply to be less predictable leading to possible mismatches in the power network. To this end, hydrogen production and storage can provide a solution by increasing flexibility within the system. Stored hydrogen can either be converted back to electricity or it can be used as feed-stock for industry, heating for built environment, and as fuel for vehicles. This research examines the optimal strategies for operating integrated energy systems consisting of renewable energy production and hydrogen storage. Using Markov decision process theory, we construct optimal policies for day-to-day decisions on how much energy to store as hydrogen, or buy from or sell to the electricity market, and on how much hydrogen to sell for use as gas. We pay special emphasis to practical settings, such as contractually binding power purchase agreements, varying electricity prices, different distribution channels, green hydrogen offtake agreements, and hydrogen market price uncertainties. Extensive experiments and analysis are performed in the context of Northern Netherlands where Europe's first Hydrogen Valley is being formed. Results show that substantial gains in operational revenues of up to 51\% are possible by introducing hydrogen storage units and competitive hydrogen market-prices. This amounts to a \euro 126,000 increase in revenues per turbine per year for a 4.5 MW wind turbine. Moreover, our results indicate that hydrogen offtake agreements will be crucial in keeping the energy transition on track.
\end{abstract}

\begin{keyword}
\small
    Green Hydrogen \sep Wind Energy \sep Markov Decision Process \sep Production Control \sep Inventory Management \sep Power Purchase Agreements \sep Market Price Uncertainty \sep Hydrogen Offtake Agreements \\[0.2cm]
    \textit{Abbreviations:} HES: Hydrogen Energy Storage; GHP: Green Hydrogen Plant; PPA: Power Purchasing Agreement; MDP: Markov Decision Process
\end{keyword}
\end{frontmatter}

\date{\today}

\newcommand\independent{\protect\mathpalette{\protect\independenT}{\perp}}
\def\independenT#1#2{\mathrel{\rlap{$#1#2$}\mkern2mu{#1#2}}}

\newcommand\nppa{\ensuremath{n^{\text{ppa}}}}
\newcommand\qppa{\ensuremath{q^{\text{ppa}}}}
\newcommand\pppa{\ensuremath{p^{\text{ppa}}}}


\section{Introduction}

The past few decades experienced a rapid increase in wind energy production. Between 2000 and 2018, wind energy usage increased from~0.2\% to~4.8\% of the total electricity production and this is expected to increase to more than~12\% in~2040 \citep{iea}. This rapid growth is partly driven by technological improvements in the wind sector. Nowadays, turbines with a peak capacity of 12 MW are being constructed, whereas only five years ago, the absolute state-of-the-art was set by the 3.6 MW turbines in offshore wind farm Gemini, the Netherlands \citep{gemini}. As a consequence of these developments, the total levelized cost of offshore wind electricity has dropped significantly and is expected to decrease by another~60\% towards~2040 \citep{iea}. However, the increasing share of wind-based electricity puts a high burden on balancing supply and demand in the electricity network. It is challenging to make advanced electricity supply commitments due to variable wind speeds and the uncertainty of future weather conditions makes reliance on offshore wind a risky investment \citep{quin}.

\textit{Hydrogen Energy Storage (HES)} systems can supplement renewable energy sources to overcome the challenges associated with higher penetrations of wind-based electricity \citep{valverde}. During periods of oversupply, electricity can be converted into green hydrogen and be stored for later use. The hydrogen stored can be converted back into electricity in times of supply shortage. Moreover, hydrogen can be used by industry, (public) transport, and for heating in built environments \citep{staffell}. The potential of these gas-based use-cases distinguishes HES from other energy storage systems such as batteries, which only offer short-term flexibility services for the electricity network \citep{minorh2}. 

The integrated power system, which we refer to as a \textit{Green Hydrogen Plant (GHP)}, will seek for the potential benefits of HES and the techno-economical efficiency increase of offshore wind farms. Such plants that jointly operate HES and offshore wind electricity production (or another renewable source) are envisioned to take a central role in a future climate-neutral society \citep{h2_nl}. Therefore, it becomes important to analyze the operational strategies for such plants in the path towards the transformation of the global energy sector. The need for such an analysis is also supported by the European Hydrogen Valleys, such as in the Northern Netherlands \citep{h2_nl2} which this study is a part of within the HEAVENN (Hydrogen Energy Applications in Valley Environments for Northern Netherlands).


Various factors affect the behavior of GHP operators. First, the distribution channels of green hydrogen (e.g., pipelines, vessels, and trucks) have a high impact on feasibility and profitability, thereby directly affecting the GHP operators' incentives. Potentially, existing natural gas infrastructure will be refurbished \citep[see, e.g.,][]{dwvm}, transportation will take place using gaseous hydrogen in tube trailers \citep{mulderoutlook}, or hydrogen will be transported by ships \citep[see, e.g., the use-case in Asia by][]{h2boot}. Second, the shape of a future (price-setting) hydrogen market will play a prominent role in the developments. For instance, hydrogen prices might be regulated centrally, locally, or even negotiated individually \citep{energyac}. In such markets, hydrogen offtake agreements will be crucial in shaping the policies of GHP operators. Third, due to the combination of high capital costs and uncertainty in future payoffs, offshore wind farms are managed using so-called Power Purchase Agreements (PPA) \citep{jenkins}. A PPA dictates that the seller (i.e., the offshore wind farm) delivers predefined amounts of renewable electricity at fixed times and fixed prices to the contractual buyer \citep[see, e.g.,][]{bolinger}. The price, size, and timings dictated in the PPA directly affect the GHP operator's behavior.

In this paper, we study the optimal control strategy of a GHP operator, that is, a renewable energy producer who owns a HES and whose wind farm is managed under both a PPA and hydrogen offtake agreements. The renewable energy producer aims to maximize its profit by deciding when and how much electricity to sell to the PPA and how much to buy from or sell to the electricity market. Moreover, the operator controls the hydrogen inventory level of the storage facility. This implies that the renewable energy producer has to decide how much electricity and hydrogen to sell, buy or store throughout the year. The control strategy considered is dynamic, that is, the decisions may depend on the observed levels of renewable energy production, the current electricity and hydrogen prices, and the amount of hydrogen in the storage facility. Moreover, the optimal decisions depend on the rules set by the PPA, hydrogen offtake agreements, and potential hydrogen distribution channels.

We formulate this stochastic optimization problem as a Markov decision process (MDP) and derive optimal policies by using backward dynamic programming. Thus, we obtain an optimal control policy for a profit-maximizing GHP operator providing insights on how the operator can interact with potential hydrogen distribution channels and the electricity market. We consider two different policies on how hydrogen as gas can be sold: First, we consider complete freedom in determining the amount of hydrogen to be sold subject to the uncertain market price. Second, we consider a contract structure with fixed quantities at fixed prices based on the hydrogen offtake agreements.

Our numerical experiments show the potential of GHPs, leading to the following managerial insights relevant for policy makers. First, in a fully functioning hydrogen market, the GHP has 8.3\% higher revenues compared to a system where power can be stored under very high conversion rates akin to a wind-to-battery system but where hydrogen is not sold as a separate product. Under the current market conditions, this increase in revenue equals \euro 22,648 per year per 4.5 MW wind turbine. Comparing the GHP to a system without any power storage options, the revenue difference increases by up to 52\%, and compared to a system with HES but without direct hydrogen distribution possibilities the revenue difference equals 33\%. Second, in the long term, under future mature hydrogen markets, dependency on the hydrogen offtake agreements becomes less prominent as this imposes additional constraints on the profit-maximizing behaviour of the GHP operator reducing the income from arbitrage opportunities. Third, hydrogen offtake agreements are nevertheless attractive if the frequency in which hydrogen should be sold is relatively high. In other words, bulk-selling hydrogen at predetermined moments in time is an attractive alternative from a profit-maximizing perspective. Fourth, although there are slight distinctions in profit between hydrogen distribution policies, we observe that if hydrogen prices are on average 11-13 \euro/MWh higher than electricity prices, the more restrictive hydrogen offtake agreements are equally profitable offsetting the current low efficiency rates of electrolyzers. This result shows that hydrogen offtake agreements will be crucial for the adoption of green hydrogen when there is no fully functioning hydrogen market.

In summary, we make the following scientific contributions. First, this paper is the first in-depth study that analyzes the optimal dynamic policies for renewable energy producers that not only interact with the electricity market but also takes into account the hydrogen market. Second, to the best of the authors' knowledge, this is the first study that thoroughly studies PPA obligations and hydrogen offtake agreements in such settings.

The remainder of this paper is structured as follows. In Section~\ref{sec:litrev}, we outline the related literature. In Section~\ref{sec:systdesc}, we provide a formal problem statement and associated mathematical model as a Markov decision process. In Section~\ref{sec:results}, we present and discuss our numerical results and provide managerial insights. In Section~\ref{sec:conc}, we conclude and provide directions for future research.


\section{Literature review}
\label{sec:litrev}

A limited number of studies relate to the fundamental problem of integrating hydrogen energy storage systems with wind power generation. In this review, we take a broader view on storage operational problems and position our work within. Studies on storage operations can be categorized by the interactions with the associated electricity markets, the renewable energy resource, and the storage form.

Interactions with the electricity market and how these operations are managed play a vital role in storage operations. \cite{kim} consider the trading of electricity on an electricity spot market and the regulating market. Similarly, \cite{lohndorf} consider bidding on a short-term intraday market and a long-term interday market. \cite{hassler} present a model for short-term trading with a time lag between trade and delivery and \cite{durante} use forecasted power output with forecast errors. \cite{qi} focus on the optimal size and sites of energy storage systems, as well as the associated topology and capacity of the transmission network under a given policy instrument. \cite{dispose} consider the influence of negative prices on energy behavior and compare energy storage with disposal. \cite{wu} focus on storage operations with limited flexibility. 

Power purchasing agreements are increasingly being adopted as private companies are becoming more and more involved in sourcing their power from renewable energy producers. The structure of PPAs can have a strong influence on the optimal behaviour of energy producers in their interactions with the electricity market. This calls for the need for more studies that investigate the influence of PPAs on the selling behaviour of renewable energy producers. In some other fields, such as in the field of maintenance, \cite{lei} and \cite{lei2} optimize predictive maintenance opportunities for wind farms managed under PPAs. They find that the optimum opportunity for a wind farm managed under a PPA differs for the same wind farm managed under a so-called `as-delivered' contract and for wind farms that are managed in isolation. \cite{davidson} evaluate how the ultimate cost of the system to the customer is impacted by the timing of payments under a Third Party Owner (TPO, managed with a PPA) contract. They conclude that especially the structure of the contract and the timing of payments have financial implications for the customer. Finally, \cite{jenkins} investigate a PPA structure and identify the relative importance of the different variables found in such an agreement. They observe that the project's financial feasibility is affected by the electric tariff, the actual plant load factor and the project cost. For a detailed analysis on the different PPAs or on the possible financing structures observed in the field of wind projects, we refer to \cite{harper}. This paper extends the studies that focus on the interaction with the electricity market by studying the role of power purchase agreements on the selling and buying behaviour of a renewable energy producer connected to a hydrogen storage system.

Among the studies that consider the interplay of generation and storage, a number of papers have a special focus on electricity production via wind energy \citep{qi, dispose,wu,bakir2021integrated}. For an overview of these papers, we refer to the work of \cite{weitzel}. In this paper, we also focus on production via wind energy and further extend this with a hydrogen storage facility. The studies of \cite{zhou, halman, hassler, choi, durante, harsha} investigate a system where a wind farm is connected to a battery system instead. Although the conversion efficiencies of batteries are higher than those of hydrogen storage, batteries are seen as less sustainable due to its polluting properties \citep{minorh2}. Moreover, batteries are far less easily scaled than hydrogen storage facilities, making batteries less suitable for large scale storage \citep{battery}. Finally, as hydrogen itself can be used in for example the industry, mobility and the built environment, energy storage in the form of hydrogen provides important additional opportunities for renewable energy producers.

The hydrogen economy is foreseen to have an important role in the global energy system of the future. As more and more countries consider hydrogen to reach their climate goals \citep{ieah2}, we notice an increase in area-specific case studies. \cite{valverde} model the performance of wind-hydrogen energy systems in Scotland and UK. \cite{hajim} optimize the size of hydrogen production plants in Ontario (Canada). \cite{gutierrez} study the feasibility of hydrogen for energy management of a wind farm Spain and \cite{gonzalez}, investigate the viability of hydrogen production using wind power that cannot easily be accommodated on the system in Ireland.
 
Hydrogen storage distinguishes itself from other storage techniques in, for example, efficiency, dis(charging) time and storage requirements. For an overview of hydrogen energy technologies (production, storage, distribution and utilization) we refer to \cite{sherif} and \cite{deshmukh}. Recently, studies were conducted to analyse the role of hydrogen as a means to provide flexibility to the power market. In the work of \cite{mirzaei}, HES is used to store excess wind power and plays a role in balancing power production and consumption. Similarly, \cite{gutierrez} propose hydrogen as a means of energy management. The installation will store the surplus energy and return to electricity to the grid during peak hours. \cite{gonzalez} investigates the role of hydrogen in enabling a large increase in wind energy and \cite{hajim} considers hydrogen energy storage to manage electricity grid constraints. As hydrogen itself can be source of energy for use in different sectors next to providing flexibility to the power market, strategies for managing these systems should be given special attention. Therefore, in this study we recognize that the renewable energy producer may also offer its hydrogen to the market next to providing power to the electricity market. This paper further distinguishes itself from related literature by considering this hydrogen market opportunities with hydrogen offtake agreements and market price.
 
In summary, this paper presents important contributions to the literature by 1) providing a first thorough analysis for the optimal strategies for renewable energy providers working under power purchasing agreements with hydrogen energy storage, and 2) developing the first joint models and optimal policies for integrated wind-power hydrogen systems that jointly interact with the electricity and the hydrogen market with market price and hydrogen offtake agreements.


\section{System Description}
\label{sec:systdesc}

We consider a single Green Hydrogen Plant (GHP) operator that is the owner of a renewable energy plant with a co-located hydrogen energy storage (HES) facility. Before describing our system in detail, we first give a brief overview of its essential components. In the remainder of this study, we refer to the renewable energy plant as an offshore wind farm, however, without structural adaptations, the system can also be thought of as any other renewable energy plant such as a solar farm. In the remainder of this section, we first give a brief problem narrative with a schematic overview of our considered system. Then, we introduce the global notation of the system, followed by formulation of the problem as a Markov decision problem. In our study, we adhere to the notation and unified framework for stochastic optimization as described by \cite{powell2019unified}.

\subsection{Problem Narrative}

The renewable energy producer manages the offshore wind farm and the HES. The HES consists of an electrolyzer to convert electricity to green hydrogen, a storage facility to store hydrogen, and a fuel cell to convert green hydrogen to electricity. The entire system is subject to a base-load PPA that describes when electricity targets have to be met. The producer is connected to the grid, via which the operator interacts with the electricity market and over which the electricity as dictated by the PPA is transmitted. The GHP operator faces stochastic electricity prices on the market. 

Next to interacting with the electricity market, and only using the HES to anticipate to price fluctuations in the electricity market, hydrogen can be sold as a gas under either a `free' or a `fixed' policy. The `free' policy assumes stochastic market prices and an unrestricted decision on when and how much hydrogen will be sold. The `fixed' policy assumes contractually fixed shipments of hydrogen will be sold under Hydrogen Offtake Agreements. From a modeling perspective, we will only consider the `free' policy while formulating the MDP, and remark the changes required to obtain the `fixed' policy to incorporate Hydrogen Offtake Agreements thereafter. 

\begin{figure} [hbt!]
    \centering
    \includegraphics[width = 12 cm]{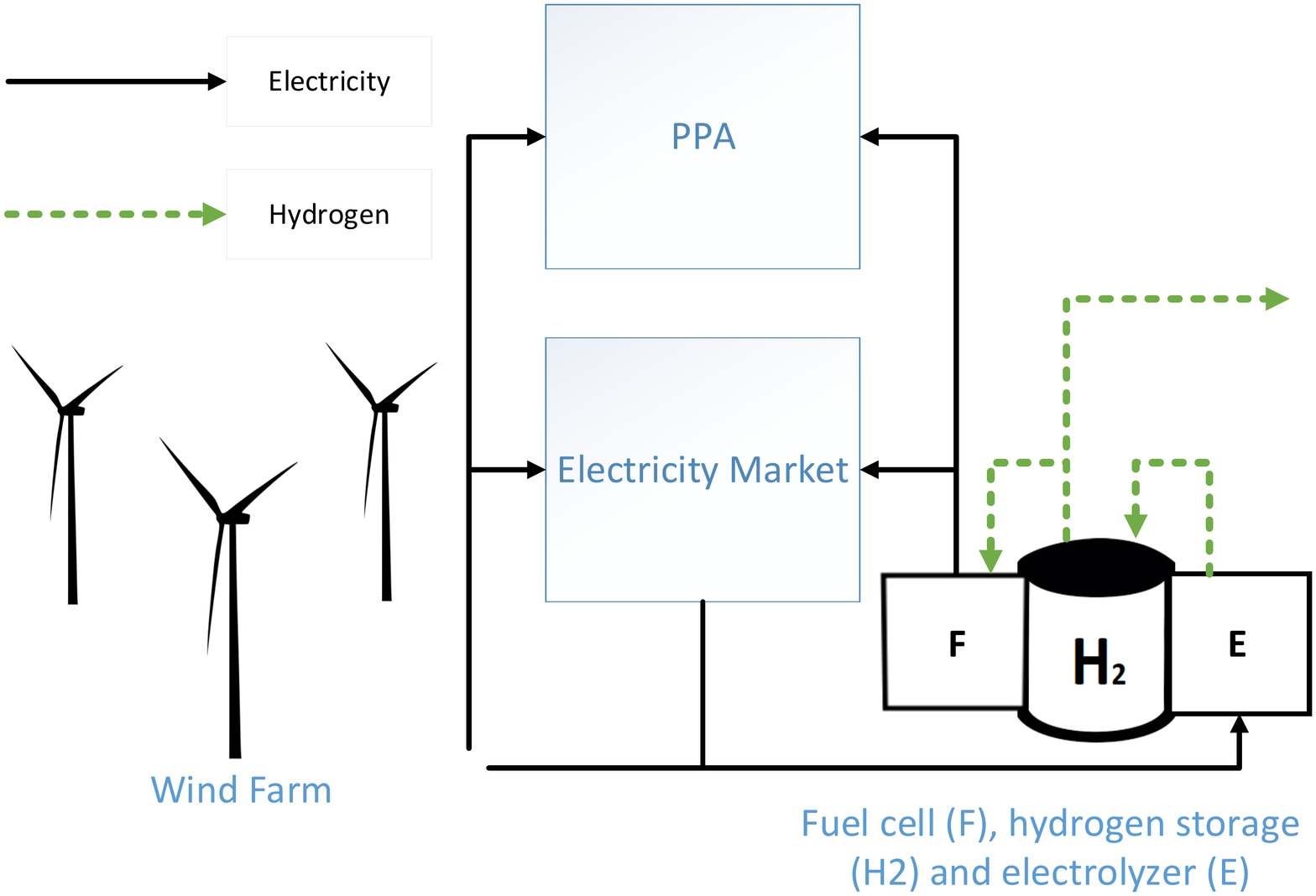}
    \caption{System overview}
    \label{fig:plaatje}
\end{figure}

Figure~\ref{fig:plaatje} gives a schematic overview of the system. In every period, the total wind energy production and the current electricity and hydrogen market prices are observed and the operator needs to decide 1) how much electricity to buy from or sell to the market, 2) how much hydrogen to sell, and 3) how much electricity to sell to fulfill the PPA requirements. As we aim to characterize the long-term optimal strategy of the GHP owner, we assume these decisions are taken simultaneously, and thereby assume that intra-day fluctuations are handled separately from our system.

\subsection{Global System Parameters}

The GHP operator manages an offshore wind farm with a fixed number of wind turbines. We consider a planning horizon $\mathcal T = \{1, \ldots, T\}$ of one year, as it is likely that system parameters such as the PPA restrictions are updated every year to account for latest market price developments. The operator faces a PPA that dictates the GHP to sell a total of $\qppa$ electricity units for a fixed price $\pppa$ every interval of $\nppa$ periods. If the PPA target is not met, a penalty $c^{\text{penalty}}$ is incurred for each unit of electricity that is short.

The HES consists of three parts. First, the electrolyzer uses electricity to convert water into hydrogen and oxygen, with an efficiency equal to $0 \leq \alpha^{\text{e}} \leq 1$ and a maximum converting capacity of $k^{\text{e}}$ (expressed in units electricity per period). Second, the produced hydrogen is transmitted to the storage facility with capacity~$Q$. Third, hydrogen can be converted back into electricity by using a fuel cell with a maximum period capacity~$k^{\text{f}}$ (expressed in units of electricity per period) and conversion efficiency equal to $0 \leq \alpha^{\text{f}} \leq 1$. Furthermore, the total amount of electricity that can be bought from or sold to the market per period is restricted by the transmission capacity $k^{\text{c}}$, and the total amount of hydrogen that can be sold to the market per period is restricted by~$k^{\text{h}}$.


\subsection{Notation used in the Markov Decision Process}

An overview of all the notation used is given in Table~\ref{tab:notation}.

\begin{table} 
    \centering 
    \renewcommand*{\arraystretch}{0.9}
    \caption{Overview of notation used}

    \label{tab:notation}
    \begin{tabular}{p{4cm}|p{12cm}} 
    \toprule
    \multicolumn{2}{c}{\textbf{Sets}}\\
    \midrule
    $\mathcal T$ & Number of periods, $\mathcal T$ = \{1,2, \ldots, T\}\\
    \midrule
    \multicolumn{2}{c}{\textbf{System parameters}}\\
    \midrule
    $\nppa$ & Number of periods between subsequent PPA targets\\
    $\qppa$ & Amount of electricity to sell to PPA per $\nppa$ periods\\
    $\pppa$ & Fixed electricity price according to PPA\\
    $\alpha^\text{e}$ & Conversion efficiency of electrolysis \\
    $\alpha^\text{f}$ & Conversion efficiency of fuel cell \\
    $\alpha$ & Round-trip conversion efficiency \\
    $k^\text{e}$ & Maximum amount of electricity that can be converted to hydrogen each period \\
    $k^\text{f}$ & Maximum amount of hydrogen that can be converted to electricity each period \\
    $k^\text{c}$ & Maximum transmission capacity each period \\
    $k^\text{h}$ & Maximum amount of hydrogen that can be sold per period \\
    $\bar{p}$ & Premium on energy price when buying from market \\ 
    \midrule
    \multicolumn{2}{c}{\textbf{State variables}} \\ \midrule
    $\mathcal S$    & State space\\
    $S_t$           & State at period $t$\\
    $p^\text{e}_t$         & Electricity selling price in period $t$ \\
    $p^\text{h}_t$         & Hydrogen selling price in period $t$ \\
    $y_t$           & Wind-energy production in period $t$ \\
    $I_t$           & Inventory level at the start of period $t$ \\
    $v_t$           & Amount of electricity not yet fulfilled according to PPA \\
    $t$             & Period \\
    $\bar{p}$       & Fixed price markup for buying from electricity market \\
    \midrule
    \multicolumn{2}{c}{\textbf{Decision variables}} \\ \midrule 
    $\mathcal{X}(s_t)$  & Space of decision variables\\
    $x_t(S_t)$          & Decision in period $t$\\
    $x_t^\text{sell}$   & Amount of electricity sold to the market \\
    $x_t^\text{buy}$    & Amount of electricity bought from market \\
    $x_t^\text{ppa}$    & Amount of electricity sold to PPA \\
    $x_t^\text{h}$      & Amount of hydrogen sold as gas \\
    $x_t^\text{in}$     & Amount of electricity converted to hydrogen after market and PPA interaction \\ 
    $x_t^\text{out}$    & Amount of hydrogen converted to electricity after market and PPA interaction \\
    $R(S_t, x_t)$       & Direct reward of decision $x_t$\\
    \midrule
    \multicolumn{2}{c}{\textbf{Exogenous information}} \\ \midrule 
    $W_{t+1}(S_t)$ & Exogenous information variable \\
    $\theta^\text{e}$ & Parameter in front of the AR(1) term, electricity selling price process\\ 
    $\theta^\text{h}$ & Parameter in front of the AR(1) term, hydrogen selling price process \\
    $\epsilon^\text{e}$ & Noise of AR electricity selling price process with standard deviation $\sigma^\text{e}$ \\
    $\epsilon^\text{h}$ & Noise of AR hydrogen selling price process with standard deviation $\sigma^\text{h}$ \\
    $\mu^{\text{e}}$ & Constant in the electricity selling price process \\
    $\mu^{\text{h}}$ & Constant in the hydrogen selling price process \\
    \midrule
    \multicolumn{2}{c}{\textbf{Transition function}} \\
    \midrule 
    $S^M(S_t, x_t, W_{t+1})$ & Transition function \\
    $S^*_t = (t^*, p^\text{e}_{t^*}, p^\text{h}_{t^*}, I_{t^*}, v_{t^*})$ & Post-decision state \\
    \midrule
    \multicolumn{2}{c}{\textbf{Objective function}} \\ \midrule  
    $\Pi$       & Set of policies \\
    $X^{\Pi}$   & Decision rule, $\pi \in \Pi$.\\
    \bottomrule
    \end{tabular}

\end{table}

\subsection{Markov Decision Process}

We now formulate our problem as a Markov Decision Process (MDP). We describe the state variables, the decision variables, the exogenous information function, and the transition function. The notation used is based on the unified framework for stochastic optimization by \cite{powell2019unified}. Finally, we present the objective of the GHP operator. For readability and without loss of generality, we assume $\alpha^\text{e} = \alpha^\text{h}$, so that round-trip conversion is modelled as-if it incurs after electrolysis. We further assume that appropriate discretizations have been made of the price and production processes, so that all the introduced variables are measured in units of energy. We reflect upon that at the end of this section.

\subsubsection{State Variables}

We define $S_t$, $t \in \mathcal{T}$ as the state observed at period~$t$. We assume that the decision epochs coincide with the period set~$\mathcal{T}$. A state of the system $S_t \in \mathcal S$ is then formally described as $S_t = (t, p_t^\text{e}, p_t^{\text{h}}, y_t, I_t, v_t)$. Here, $p_t^{\text{e}}$ denotes the electricity price in period $t$, $p_t^{\text{h}}$ the hydrogen market price in period~$t$, $y_t$ the wind energy production in period~$t$, $I_t$ the inventory level at the start of period~$t$, and~$v_t$ denotes the amount of electricity that still has to be sold according to the PPA.

\subsubsection{Decision Variables}

The decision variable $x_t(S_t) \in \mathcal X(S_t)$ is described by a vector of four variables that indicate how much electricity is sold to the market, how much electricity is bought from the market, how much electricity is sold according to the PPA (in addition to the market interaction), and how much hydrogen is sold. We write $x_t = (x^{\text{sell}}_t, x^{\text{buy}}_t, x^{\text{PPA}}_t, x^{\text{h}}_t)$. 

The action space $X(S_t)$ is restricted by the constraints of the system and depends on state~$S_t$. For this purpose, let $x^{\text{in}}_t$ denote the amount of energy units placed into storage after interacting with the electricity market and the PPA, and let $x^{\text{out}}$ denote the amount of energy units moved out of the storage after interacting with the electricity market and the PPA. 
The feasible action space has to adhere the following constraints:
\begin{enumerate}
    \item We can sell to or buy from the market but not both, hence $x^{\text{sell}}_t x^{\text{buy}}_t = 0$.
    
    \item If $x^{\text{sell}}_t > 0$ and $x^{\text{buy}}_t = 0$, then the following constraints must hold:
    \begin{enumerate}
        \item The total amount of energy sold should satisfy the transmission capacity: $x^{\text{sell}}_t + x^{\text{PPA}}_t \leq k^{\text{c}}$,
        \item As there is no electricity bought from the market, we convert all the production that is left after market interaction: $x^{\text{in}}_t = \max(0, y_t - x^{\text{sell}}_t - x^{\text{PPA}}_t) \alpha$, where $\alpha = \alpha^\text{e}\alpha^\text{f}$,
        \item For the same reason, the amount of energy leaving the storage equals the production in excess of the electricity sold to the market: $x^{\text{out}}_t = \max(0, x^{\text{sell}}_t + x^{\text{PPA}}_t - y_t)$.
    \end{enumerate}
    
    \item If $x^{\text{sell}}_t = 0$ and $x^{\text{buy}}_t > 0$, then the following constraints hold:
    \begin{enumerate}
        \item The total amount of energy bought should fit in storage: $x^{\text{buy}}_t  < \min \{ (Q - I_t)/\alpha, k^{\text{c}}\}$,
        \item We cannot simultaneously buy and sell, thus we convert the electricity bought from the market and the production not used for satisfying the PPA: $x^{\text{in}}_t =  \alpha x^{\text{buy}}_t + \alpha\max\{0, Y_t - x^{\text{PPA}}_t\}$,
        \item For the same reason, energy that leaves the storage is only used for meeting PPA obligations: $x^{\text{out}}_t = \max(0, x^{\text{PPA}}_t - Y_t)$.
    \end{enumerate}
    
    \item Electrolyzer, fuel cell, and inventory capacity should be respected: $x^{\text{in}}_t \leq k^\text{e}$, $x^{\text{out}}_t \leq k^\text{f}$, and $0 \leq I_t + x^\text{in}_t - x^\text{out}_t \leq Q$
    
    \item The maximum amount of hydrogen that can be sold equals: $x^{\text{h}} = \min \{I_t +  x^{\text{in}}_t - x^{\text{out}}_t, k^\text{h}\}$.
\end{enumerate}

The direct reward of action $x_t$ equals $R(S_t, x_t) = p_t^{\text{e}}x_t^{\text{sell}} - (p_t^\text{e} + \bar{p})x_t^\text{buy} + R^{\text{PPA}}(S_t, x_t) + p^{h}_tx^{\text{h}}_t$ . The first term is the reward from selling electricity to the market, the second term is the reward from buying electricity from the market, the third term ($R^\text{PPA}(S_t, x_t)$) denotes the reward from meeting PPA obligations, and the fourth term denotes the reward from selling hydrogen to the hydrogen market. We define $R^\text{PPA}(S_t, x_t) = p^{\text{PPA}}x^\text{PPA}_t - c^{\text{penalty}}\max \{0, v_t - x^{\text{PPA}}_t \} 1_{[t \mod n^{\text{PPA}}] = 0}$. That is, we obtain revenues for selling electricity for the fixed PPA price and we pay the penalty for not meeting the PPA target if period $t$ coincided with a PPA deadline. The latter is modelled via the indicator function $1_{[t \mod n^{\text{PPA}}] = 0}$ equalling 1 if $[t \mod n^{\text{PPA}}] = 0$ and 0 otherwise. Note, the impact of the fixed hydrogen selling policy will be discussed at the end of this section.

\subsubsection{Exogenous Information}

After each decision point~$t$, we observe new market electricity prices, new hydrogen prices, and wind-energy production. The exogenous information variable is denoted by $W_{t+1}(S_t)$ and does not depend on the action taken in period~$t$. The market electricity prices are stochastic and follow an AR(1) process  $p_{t+1}^\text{e} = \mu^{\text{e}} + \theta^\text{e} p_{t}^\text{e} + \epsilon^\text{e}, \epsilon^\text{e} \sim \mathcal{N}(0,\sigma^\text{e})$, as do the hydrogen market prices $p^\text{h}_{t+1} = \mu^{\text{h}} + \theta^\text{h} p_{t}^h + \epsilon^\text{h}, \ \epsilon^\text{h} \sim \mathcal{N}(0,\sigma^\text{h})$. The production~$I_{t+1}$ follows a Weibull distribution with period dependent shape and scale parameters. 

\subsubsection{Transition Function}

The transition function $S^M(S_t, x_t, W_{t+1}) = S_{t+1}$ describes how we transition towards the state in period~$t+1$. It first applies the feasible action~$x_t$ to reach a post-decision state~$S^*_t$, after which the exogenous information determines the transition from~$S^*_t$ into~$S_{t+1}$. 

Let $S^*_{t} = ({t^*}, p_{t^*}^\text{e}, p_{t^*}^\text{h}, I_{t^*}, v_{t^*})$, where ${t^*} = t$, $p_{t^*}^\text{e} = p_t^\text{e}$, and $p_{t^*}^\text{h} = p_t^\text{h}$. 
\begin{enumerate}
    \item The post-decision inventory level $I_{t^*} = I_t + x^\text{in}_t - x^\text{out}_t - x^\text{h}_t$. 
    \item The post-decision target $v_{t^*} = v_t - x^\text{ppa}_t$ in case $t \mod n^{\text{PPA}} \neq 0$, else $v_t^{*} = q^\text{PPA}$.
\end{enumerate}
Then, the transition to $S_{t+1}$ follows trivially by including the new information from $W_{t+1}$. That is, $S_{t+1} = (t+1, p^\text{e}_{t+1}, p^\text{h}_{t+1}, I_{t+1}, v_{t + 1})$ with $I_{t+1} = I_{t^*}$ and $v_{t+1} = v_{t^*}$.

\subsubsection{Objective Function}

The objective of a profit-maximizing GHP operator is then given by a decision policy $\pi \in \Pi$ and a decision rule $X^\pi: {\cal S}_t \rightarrow {\cal X(S}_t)$. Thus, we write $x_t = X^\pi(S_t)$ as the decision $x_t$ under decision policy $\pi$. The objective of the GHP operator is then to maximize its total expected profit:
\begin{align}
    \max_{\pi \in \Pi} \mathbb{E}\Big[\sum_{t \in \cal T} R(S_t, X^\pi(S_t))\mid S_0\Big], \label{eq:bell}
\end{align}
where $S_{t+1} = S^M(S_t, X^\pi(S_t), W_{t+1})$, and $S_0$ denotes the state at the start of the year. 

\subsection{Solution Approach, Discretization, and Hydrogen Policies}

We solve Bellman equation~\eqref{eq:bell} to optimality by using backward dynamic programming on a discretized set of electricity and hydrogen prices, production levels, inventory levels, and PPA target levels. The electricity production is obtained by considering Weibull distributed wind speeds per period, and converting them via a power curve of a wind-turbine to associated production levels. The amount of considered production levels~$L^\text{y}$, which is a parameter of our discretized model, then determines the equivalent electricity amount of a one energy unit in our system. We provide exact details on the wind turbine power curve modeling in the numerical results. Besides, we discretize the hydrogen and electricity market processes in~$L^\text{e}$ and~$L^\text{h}$ distinct price levels, respectively. 

For the conversion efficiency, we assume that the fuel cell conversion loss is incurred when converting electricity to hydrogen, and we accommodate this in the hydrogen price accordingly such that conversion losses are presented within. This implies that selling one energy unit (in our model) as hydrogen has a higher equivalent electricity amount than the sold electricity with no  structural differences in the results.

The MDP formulation allows hydrogen being sold in every period subject to market prices. We extend our above policy where we restrict selling hydrogen only once every $n^\text{h}$ periods while respecting the system constraints on capacity and distribution. This extension is referred to as Policy $B(n^\text{h})$ in our numerical results. Further, we consider a policy $C(n^\text{h}, \bar{p}^\text{h}, Q^\text{h})$ that is  subject to $Q^\text{h}$ fixed amounts of hydrogen to be sold \textit{on} every $n^\text{h}$-th day of the planning horizon at price $\bar{p}^h$. Our MDP can be straightforwardly adapted to account for both policies. Namely, we restrict the action $x^\text{h}_t$ and the reward function $R(S_t, x_t)$ so that is in line with the above mentioned policies. Note that, if hydrogen prices are fixed, the AR process of hydrogen prices becomes redundant. Finally, if there is not enough hydrogen on stock when it needs to be sold, we assume shortages are penalized similar to the PPA shortage penalty price.


\section{Numerical Analysis}
\label{sec:results}

In this numerical section, we evaluate the economic viability of GHP. We compare different hydrogen settings, which we introduce in Section~\ref{s:setting_desc}. We continue by describing the system parameters and how the price and production processes are obtained in Section~\ref{s:basecase_d}. Next, we present the comparison of the different settings on various Key Performance Indicators (KPIs) in Section~\ref{sec:opt_pol_per}. We continue with analyzing the profitability of our settings on future hydrogen markets in Section~\ref{s:sens_hydrogenprice}. We end with a discussion of the obtained results. 

\subsection{Setting Descriptions}
\label{s:setting_desc}

Developments in the hydrogen economy can take different forms. Therefore, the viability of GHP is evaluated under various settings described in Table~\ref{tab:h2settings}. These settings differ in the way hydrogen can be traded as gas, where we distinguish between `free' and `fixed' policies. The `free' policies, studied in settings~$A$ and~$B(n^{\text{h}})$, assume market prices and a free decision on how much hydrogen can be sold. Under setting~$A$, hydrogen can be sold every period, while under setting~$B$ this is restricted to once every $n^{\text{h}}$ periods which is a possible restriction within fixed shipment settings where hydrogen is transported via trailers. The `fixed' policy, named as setting~$C(n^\text{h}, \bar{p}^\text{h}, Q^\text{h})$, assumes that contractually fixed shipments of hydrogen of size $Q^{\text{h}}$ will be sold every $n^{\text{h}}$ periods for a fixed price $\bar{p}^{\text{h}}$. Remaining settings in Table~\ref{tab:h2settings} further restrict the options of the GHP operator. Under setting $D(s), s \in\{H2, B\}$ electricity is stored using HES  without the possibility to sell hydrogen with ($s = H2$) or ($s = B$). We set that~$D(H2)$ and~$D(B)$ differ in terms of round trip efficiency with~$D(B)$ representing a very high hypothetical conversion rate. Note that although we start with two different levels we further enrich our analysis for intermediary levels. Finally, setting~$E$ considers the setting where we do not have any form of storage. Note that setting~$D(s)$ can be obtained as setting $C(0, 0, 0)$, while setting~$E$ can be obtained $C(0, 0, 0)$ with no electrolyzer and fuel cell capacity.

\begin{table}[t!]
    \centering
    
    \caption{Hydrogen market trading structures}
    \label{tab:h2settings}
    
    \begin{tabular}{lllll} \toprule
    {Setting} & {Amount of hydrogen sold} & {Price} & {Periods of selling}  & Constraints\\ \midrule
    $A$    & Variable & Market price & Every period & - \\
    $B(n^\text{h})$ & Variable & Market price& Once every $n^\text{h}$ periods & - \\ 
    $C(n^\text{h}, \bar{p}^\text{h}, Q^\text{h})$ & $Q^\text{h}$ & $\bar{p}^\text{h}$ & Once every $n^\text{h}$ periods & - \\
    $D(s)$ & 0 & 0 & No hydrogen being sold  & - \\
    $E$ & 0 & 0 & No hydrogen being sold & No HES \\ \bottomrule 
    \end{tabular}
\end{table}

\subsection{Base-case System Description}
\label{s:basecase_d}

As a benchmark, we base our evaluation on an Enercon E112 turbine, located in the Netherlands. We consider a time horizon of a single year, thus every period~$t$ in our model corresponds to a single day. The turbine has a peak capacity of 4.5 MW and is connected to a local electricity grid through a cable with a maximum capacity of 5 MW (120 MWh per day). The process is discretized such that one production or energy unit refers to 5.7 MWh. A hydrogen storage with a capacity of 1100 MWh (200 production units) is connected to a 5 MW electrolyzer and a 5 MW stack of fuel cells. For this benchmark case, we assume round-trip efficiencies of 0.5 ($s = H2$) and 0.9 ($s = B$). Then, coupled fuel cell and the electrolyzer are assumed to have an in-efficiency of~$\sqrt{0.5}$ or~$\sqrt{0.1}$. The (relative) values of the considered generation, transmission, storage and conversion capacities are consistent with those in \cite{zhou, fokkema, steilen}.

We assume that daily wind energy production levels are stochastic. For each day, the daily wind speed levels between 2000 and 2020 are obtained from CBS \citep{data:knmi}. Similar to \cite{windaverage}, we consider the average of locations in coastal regions where most wind turbines are located, namely De Kooy, Lelystad, Leeuwarden, Lauwersoog and Wilhelminadorp. For each month, the daily observations are fitted to a Weibull distribution, which is commonly used in modeling wind speed \citep{windweibull1, windweibull2, uit2019evaluating, schrotenboer2020mixed}. Accordingly, daily wind speed levels are represented by shape and scale parameters~$\alpha_t$ and~$\beta_t$. The wind speed data was recorded at 10 meters above sea level \citep{data:knmi}, and subsequently adapted to the axle height of the Enercon E112 turbine that equals 125 meters.

To transform wind speed levels to wind energy, we follow the work by \cite{wind_production}. The power output~$P_w$ from a wind turbine generator can be calculated as follows: 
\begin{center}
    $P_w = 
    \begin{cases}  
        0 & \text{ if } V < V_{ci}, \\ 
        aV^3 -bP_r & \text{ if } V_{ci} < V < V_r, \\
        P_r & \text{ if } V_r < V < V_{co}, \\
        0 & \text{ otherwise,}
    \end{cases}$
\end{center}
where $a = P_r/(V_r^3 - V_{ci}^3), b = V_{ci}^3/(V_r^3 - V_{ci}^3),$ $P_r$ is the rated power and $V_{ci}, V_{co}$ and $V_r$ are the cut-in, cut-out and rated speed of the wind turbine. Regarding the benchmark, an Enercon E112 turbine has a cut-in speed of 3 m/s a cut-out speed of 25 m/s and a rated speed of 13 m/s \citep{wind_turbine}.  

We assume that electricity prices exhibit similar behaviour to wholesale day-ahead electricity prices.  Day-ahead hourly wholesale electricity prices (in euro/MWh) in the Netherlands between 2015 and 2019 follows the study by \cite{fokkema} based on the ENTSOE Transparency Platform \citep{data:entsoe}. In the considered period, the price levels varied from 15.4 \euro / MWh to 89.0 \euro / Mwh with a mean of 40.0 \euro / MWh. These are fit to an AR(1)-process with different seasonality filters based on the standard error, which resulted in the conclusion that no clear seasonality effect is observed in our data. Therefore, we choose our AR process parameters equal for each time period, that is, $\mu^{\text{e}}$, $\theta^\text{e}$ and $\sigma^{\text{e}}$ are set equal to 0.873, 5.23 and 5.551, respectively. Thus, $p_t$ is measured in \euro/MWh.

Within the free settings ($A$ and $B$) we consider a hydrogen price that is competitive with the electricity price where $\theta^\text{h}$ and $\sigma^{\text{h}}$ are in line with $\theta^\text{e}$ and $\sigma^{\text{e}}$, respectively. Nevertheless, we observe from practice that the hydrogen price varied in 2019 between \euro 46.0 /MWh and \euro 180.0 /MWh, which is higher than the electricity price in that period \citep{bloomberg}. This implies that our results are conservative with regards to potential hydrogen profits. We further supplement our analysis with less conservative hydrogen market settings in Section~\ref{s:sens_hydrogenprice}.

For setting~$B$, we evaluate for~$B(7)$ and~$B(14)$, such that hydrogen can be sold once every week or once every two weeks, respectively. For the fixed setting~$C$ we consider four different cases. Based on the average amount of hydrogen that is being sold in the experiments of setting~$A$, we set $C(1, 35, 3)$, in which a fixed amount of 3 production units (17 MWh) is to be transmitted every day, for a fixed price of \euro 35 /MWh. We further form consider settings $C(1, 35, 4)$, $C(7, 35, 20)$ and $C(14, 35, 30)$ in a similar manner. When a hydrogen agreement is not met, a penalty equal to \euro 200 per MWh must be paid. This assures that actions are taken in such a way that shortages are most unlikely to occur.  

The GHP is managed with a baseload-PPA. A target (benchmarked at the lower side) is set to 5 production units (28.5 MWh) evaluated every seven days. The wind farm owner is paid \euro 35 for every MWh that is sold to the PPA. This is slightly lower than the average day-ahead hourly electricity price between 2015 and 2020 as PPAs are often regarded as a financing instrument for the offshore wind farm owners. In the case where a target is not met, a penalty equal to \euro 200 per MWh must be paid. 

\subsection{Setting Performance on KPIs}
\label{sec:opt_pol_per}

\begin{figure}[htb]
    \centering
    \includegraphics[width=\textwidth]{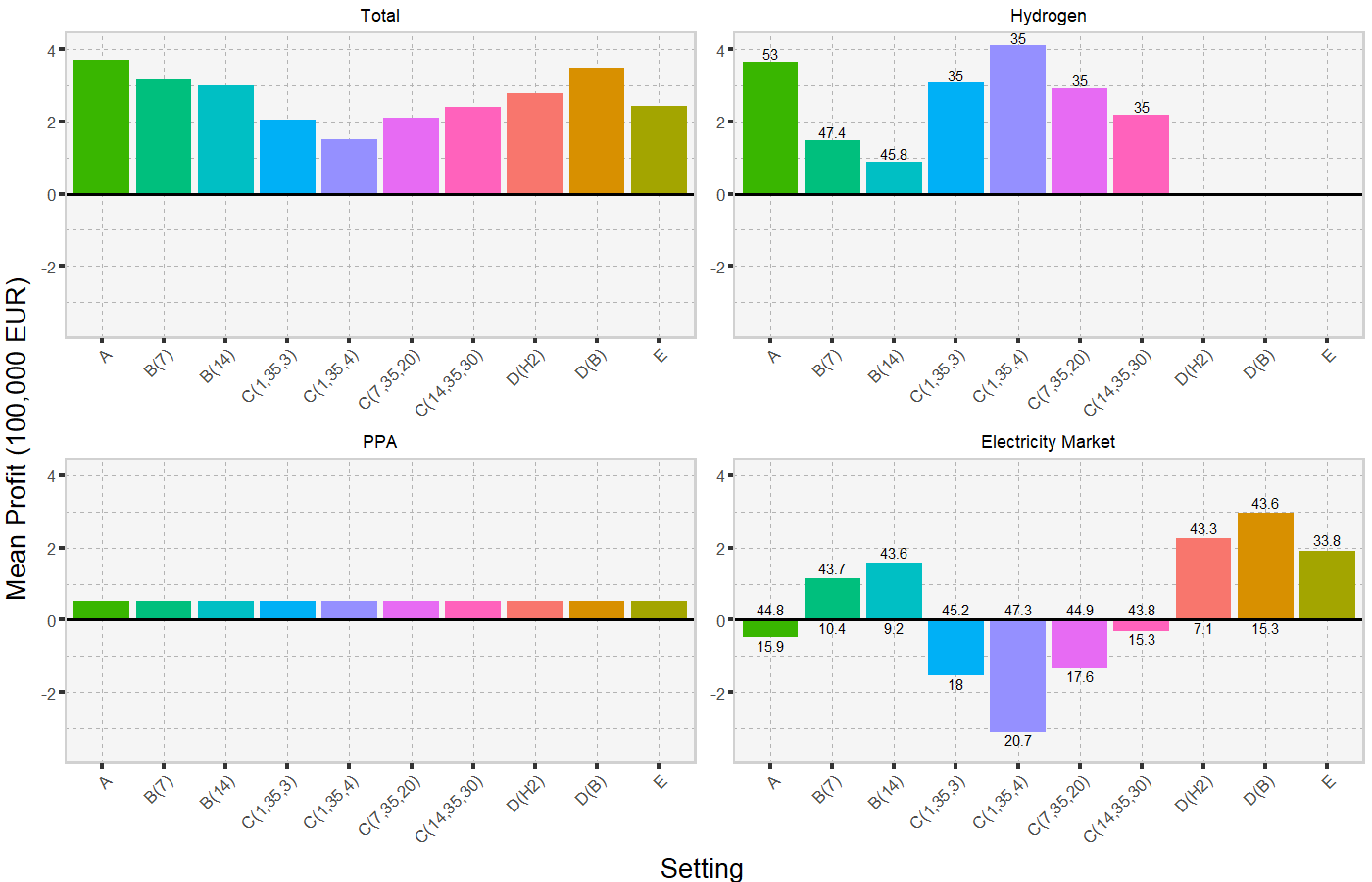}
    \caption{Total profit, hydrogen market profit, PPA profit and electricity market profit for base case}
    \label{fig:profit}
\end{figure}

We first obtained the optimal solution under each hydrogen setting setting of the base-case system. Then, we simulated the optimal setting for 10 years to obtain the relevant statistics, with in total 100,000 replications. Figure~\ref{fig:profit} shows the total profit of the considered settings (top left) and the distribution of these profits. We distinguish between profit made by selling hydrogen (top right), by satisfying the PPA (bottom left) and by trading at the electricity market (bottom right). Note that the latter can be negative as buying from the market is allowed. For the profit that is made by selling hydrogen (top right), the numbers on top of the bars indicate the average price of hydrogen that was sold in \euro/MWh. Similarly, the numbers above (below) the bars in the bottom right figure indicate the average price of electricity that was sold (bought) in~\euro/MWh. 

From Figure~\ref{fig:profit} we observe that the opportunity to sell hydrogen ($A$) increases the yearly profit, compared to the setting where we can not sell hydrogen $(D(H2))$ and when we do not have any form of energy storage~$(E)$.

We note the importance of a free hydrogen market. Comparing setting~$A$ with  setting~$B$, we observe that the current low efficiency rates of electrolyzers can be offset by this market opportunities. Before discussing this in more detail for settings with hydrogen offtake agreements, we recall our conservative hydrogen market prices and we further elaborate on that in Section~\ref{s:sens_hydrogenprice}

Considering the settings with HES, the gain is largest under setting~$A$, which allows for the most freedom to the plant operator. In terms of total yearly profit, this setting is followed by both~$B$ settings, which limit the number of periods at which hydrogen can be sold. The fixed settings~$C$, which consider a fixed hydrogen price, provide less profit than the free settings~$A$ and~$B$, but still more than setting~$E$, where we can not store any hydrogen. setting~$D(H2)$ performs worse than the free settings and most fixed settings, but better than setting $C(1, 35, 4)$. 

When observing the distribution of the profits, we see that for the optimal setting~$A$ most profit share is gained via selling hydrogen. In this setting we can exploit the fluctuations in the hydrogen market price and earn on average \euro 53.0 per MWh. This is better than the average price of the AR(1) process (\euro 40/MWh) and the price considered for the fixed settings (\euro 35/MWh). Under settings~$B(7)$ and~$B(14)$ we have the same opportunity, but the gain is smaller as we can only sell only once in a week or once in two weeks, respectively. We note that with setting $C(1, 35, 4)$ we make more profit in selling hydrogen than with setting~$A$. However, this extra profit is lost by the costs that come with buying extra electricity. It is important to add that in the case of price drops in future electricity prices due to increased production via renewable energy, hydrogen offtake agreements as in setting~$C$ may prove to be more beneficial. Compared to setting~$A$, due to the periodical restrictions, we sell relatively less hydrogen under settings~$B(7)$ and~$B(14)$. Under these settings, the profits made by trading on the electricity market are relatively large, but not large enough to compensate for the additional profit gained by selling hydrogen. We observe that under all settings the PPA is satisfied to prevent high costs.

Table~\ref{tab:kpi} shows additional key performance indicators for each of the distribution settings, where the headers indicate the yearly profit (\euro), the yearly amount of hydrogen sold (MWh), the yearly total energy loss due to conversion (MWh) and four additional indicators of the behaviour of profit-maximizing GHP operators. From left to right, those indicate the probability that, on a given day, 1) hydrogen is being sold, 2) the probability that electricity is bought from the market, 3) the probability that electricity is sold to the market and 4) the probability that electricity is transmitted to satisfy the PPA. The probabilities are calculated as the fraction of days at which the associated event occurs and as such describe the optimal long-term behaviour of the GHP operator.

\begin{table}
    \centering 
    \begin{tabular}{l|rrrrrrr}\toprule 
    setting & Profit & $H_2$ Sold &   Energy lost & $P(x^\text{h} > 0)$  & $P(x^\text{buy}_t > 0)$ &  $P(x^\text{sell}_t > 0)$ & $P(x^\text{PPA}_t > 0)$ \\ \midrule 
A & 370{,}009 & 858 & 1284 & 0.13 & 0.24 & 0.41 & 0.40 \\ 
  B(7) & 316{,}339 & 392 & 777 & 0.04 & 0.16 & 0.50 & 0.38 \\ 
  B(14) & 300{,}929 & 243 & 699 & 0.02 & 0.13 & 0.54 & 0.38 \\ 
  C(1,35,3) & 205{,}796 & 1092 & 1556 & 1.00 & 0.34 & 0.28 & 0.43 \\ 
  C(1,35,4) & 151{,}903 & 1456 & 2063 & 1.00 & 0.45 & 0.20 & 0.43 \\ 
  C(7,35,20) & 211{,}275 & 1040 & 1484 & 0.14 & 0.31 & 0.30 & 0.43 \\ 
  C(14,35,30) & 240{,}150 & 780 & 1143 & 0.07 & 0.23 & 0.37 & 0.41 \\ 
  D(H2) & 278{,}523 & 0 & 630 & 0.00 & 0.07 & 0.60 & 0.38 \\ 
  D(B) & 348{,}352 & 0 & 1911 & 0.00 & 0.15 & 0.55 & 0.37 \\ 
  E & 243{,}495 & 0 & 0 & 0.00 & 0.00 & 0.70 & 0.40 \\ 
\bottomrule 
    \end{tabular}
    
    \caption{Overview of key performance indicators for settings considered.}
    \label{tab:kpi}
\end{table}

From Table~\ref{tab:kpi}, we observe that having a storage facility increases mean profit per year from \euro 243,495 to \euro 278,523 per year ($D(H2)$ and $E$), an increase of~14.4\%. Adding the opportunity to sell hydrogen further increases the mean profit per year up~52\% to \euro 370,009 under setting~$A$. Compared to~$D(B)$, setting~$A$ increases the expected yearly profit with~6.2\%. This~6.2\% increase is even more impressive realizing that hydrogen market is modelled quite conservative, as the underlying price process is similar to the electricity market prices and setting~$D(B)$ has a hypothetically extremely high conversion rate. We note that, domain experts within our consortium foresee hydrogen offtake agreements to exhibit hydrogen prices higher than the expected electricity market prices in the near future. Main arguments to this are the immature state of green hydrogen production and that green hydrogen demand will be larger than the supply available. As this becomes an important parameter for future decisions we continue with further analysis in Section~\ref{s:sens_hydrogenprice}. 

A comparison between settings~$E$ and~$C$ shows that via the use of hydrogen offtake agreements investors can substantially increase their profits. We expect that in the short term, under a newly emerging hydrogen market economy, such hydrogen offtake agreements will play a crucial role and provide the first step towards a mature market.

Under all considered settings with HES (A-D), we notice the high amounts of losses during energy conversion. This is an interesting outcome as it implies that hydrogen is not only produced for selling in the market, but heavily so for price arbitrage opportunities. Looking into the future, HES can thus become attractive for short term flexibility mechanisms to match supply and demand in power networks, especially under local energy tariffs. Interestingly, the amount of energy lost in $D(B)$ is second-highest although it has a conversion efficiency of 0.9 compared to 0.5 of the other settings. This implies that with the lower margins necessary due to a lower energy loss per unit, the GHP operator tends to excessively buy and sell from the electricity market to make use of price differences over time. Although this might seem beneficial from a financial perspective, one should also note the sustainability view and avoid such excessive energy losses.

The fifth column of Table~\ref{tab:kpi} shows that under setting~$A$, hydrogen is on average sold on~13\% of the days. If these days are spread evenly around the year, we would sell once every 6-7 days. Under the fixed settings, we are allowed to sell hydrogen once every seven (or fourteen) days, such that $P(x^\text{h} > 0)$ is at most 0.14 (0.07). Again, according to the optimal setting, we do not sell on every period (where we are allowed) but may opt to skip some shipments for overall profitability. On the other hand, we do adhere to these shipment periods under the fixed settings~$C$, due to the relatively large penalty we face when not meeting the agreements. This also shows the need for dynamic decisions (and H2 distribution) as we propose in setting~$A$. 

When studying the probabilities to buy and sell electricity on the market (columns six and seven in Table~\ref{tab:kpi}), we see that the average probability to buy electricity is generally  higher for the fixed settings. This is natural as these settings have a strict requirement to satisfy the hydrogen demand which leads to decide on buying at a wider range of prices. Other settings generally tend to follow a similar strategy for engaging with the power market. Further, in column~5, we observe that setting~$A$, with the free hydrogen market opportunity, has an increased probability of selling hydrogen per period which reflects back as higher profits. The remaining indicator relates to the probability of selling to the PPA per period (column 8). For this we observe a similar pattern for all settings. We remark that selling to the PPA is not only done at the period at which it is due but it is spread out to the earlier period to reduce risk of not meeting PPA targets, dependent on electricity market prices. This indeed motivates the need for dynamic decision making as we propose in this study. 

\begin{figure}[ht!]
    \centering
    \includegraphics[width=10cm]{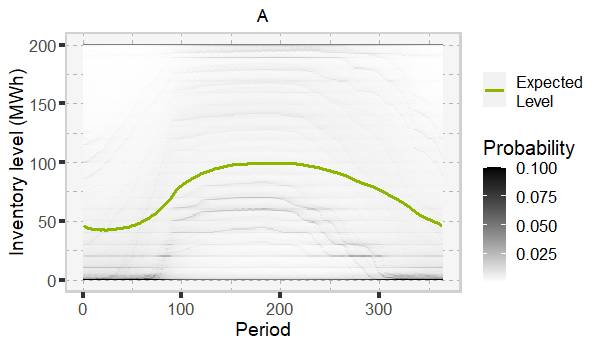}
    \caption{Heatmap of inventory level in tank over time, for setting~$A$}
    \label{fig:inventoryHeatmap}
\end{figure}

Inventory level observations have strategic importance at the long run. To see how the inventory capacity is being used throughout the year, we study Figure~\ref{fig:inventoryHeatmap} which presents the average probability to observe a particular hydrogen inventory level at each period for setting~$A$. We also show the expected hydrogen inventory level throughout the year with the bold green line. We remark that other settings with HES have a similar structure, and therefore omit these.

We notice that the expected inventory level under setting~$A$ is higher in the summer than in winter. This is caused by the fact that wind speeds (and thus electricity production) are expected to be lower in summer than in winter with higher variability. Differently, during the winter months, we observe lower variability with a higher expected mean. As power and hydrogen price processes and the power purchase agreement do not change throughout the year, more energy is kept in storage during the months where expected production as a sort of safety stock against not meeting contractual obligations. Information at this level may be important for the GHP operators as this may affect their decisions regarding storage facilities. For example, GHP operators may opt for rental storage facilities and arrange for seasonal contracts such that lower amount of storage units may be rented during low season.

\subsection{Setting Performance Under Future Hydrogen Markets}
\label{s:sens_hydrogenprice}

\begin{figure}[ht!]
    \centering
    \includegraphics[width = 1.1\textwidth]{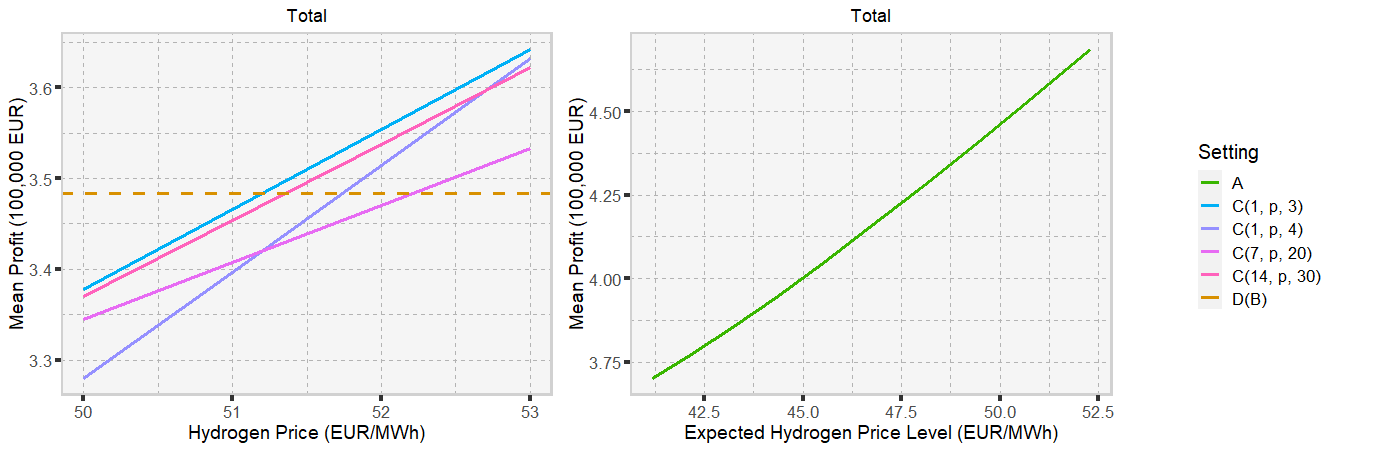}
    \caption{Impact of higher fixed hydrogen price on profit of settings with hydrogen offtake agreements, and impact of higher expected hydrogen market price (right) for setting~$A$}
    \label{fig:price_sens}
\end{figure}

Our results in the previous section showed that the GHP operator gains the highest profit margins under a mature hydrogen market (i.e., setting~$A$). However, in the near future, hydrogen offtake agreements are expected to be the norm and they will form the first step towards a mature hydrogen market. Therefore we conduct further experiments to study under which settings and market prices the hydrogen offtake agreements may be financially attractive. Thus, we investigate when the fixed hydrogen settings~$C$ become as profitable as setting~$D(B)$, where we work under a very high conversion rate with no hydrogen market. In our previous experiments we set the fixed hydrogen selling price equal to the PPA price (\euro 35). We now conduct a series of experiments to see how the~$C$ settings behave as prices approach to \euro 55/MWh. Afterward, we perform a similar analysis on setting~$A$ by structurally increasing the expected hydrogen market price.

In Figure~\ref{fig:price_sens}, we show the impact of changing the fixed hydrogen price $\bar{p}^\text{h}$ on the $C$-settings considered before (left). As a reference, we plotted the performance of setting~$D(B)$, that led to a profit of \euro 348,352, and the performance of setting~$A$ with higher mean hydrogen prices. What stands out from Figure~\ref{fig:price_sens} is that for fixed hydrogen prices only marginally higher than the expected electricity price (\euro 40 /MWh), GHP becomes as profitable as a system with very high conversion efficiencies. To be precise, the settings $C(1, 51.195, 3), C(1, 51.738, 4), C(7, 51.351, 20), C(14, 52.209, 30)$ have high enough profits to compensate for the conversion efficiency. The prices are only a 0.11-0.13 \euro/kWh higher than the expected electricity price. When looking back at the rise of offshore wind energy, which governments stimulated by guaranteeing minimum electricity prices, this difference can be reasonably mitigated by market regulations. Reflecting upon the other KPIs, we remark that next to the high gains these hydrogen offtake agreements can offer, these $C$ settings also waste less energy than the settings where intensive power market interactions take place. This is an important insight from a sustainability perspective.

\subsection{Setting Performance Under Different PPA Structures}
\label{s:sens_PPA}

\begin{figure}[ht!]
    \centering
    \includegraphics[scale = 0.6]{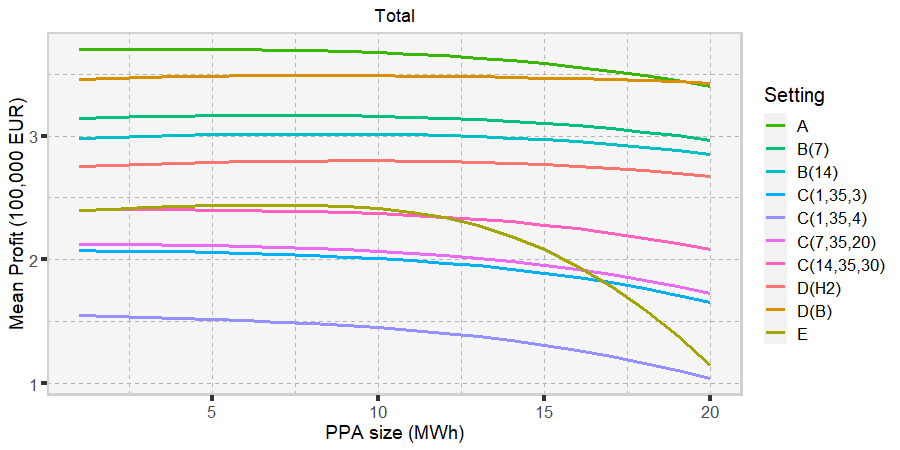}
    \caption{PPA size}
    \label{fig:ppa_totalreward}
\end{figure}

In our benchmark case we had set the PPA target to 5 production units (28.5 MWh). We now vary this target $q^\text{ppa}$ of the baseload-PPA between 1 and 20 production units. Figure~\ref{fig:ppa_totalreward} shows how the total expected reward changes with the PPA size.

We observe that at first, total expected reward increases with the target of the PPA, despite the fact that the price of the PPA (\euro 35/MWh) is lower than the expected electricity price found on the market (\euro 40/MWh) as it also acts as a safety instrument for the GHP operator. Although the PPA agreements greatly vary in practice we can foresee this lower prices due the role a PPA plays into financing an offshore wind farm; without a PPA a offshore wind farm cannot be financed, giving the energy buyer a strong negotiation position. When the PPA target is relatively small, the PPA forms a lower bound for the price found at the electricity market. However, when the size of the target further increases, the freedom of the GHP operator declines. More and more electricity has to be transmitted to prevent penalties, leaving less energy for taking advantage of fluctuating prices. 

The tipping point, at which an increase in the PPA target no longer increases the expected reward, is reached earlier for the fixed settings ($C$) than for the free settings ($A, B$), which is due to the fact that the fixed settings are already constraining the flexibility of the GHP operator. The results clearly show that given the system characteristics, negotiating right PPA terms is crucial for the GHP operator.

\subsection{Setting Performance Under Increasing Conversion Efficiency}
\label{s:sens_efficiency}

\begin{figure}[ht!]
    \centering
    \includegraphics[width=\textwidth]{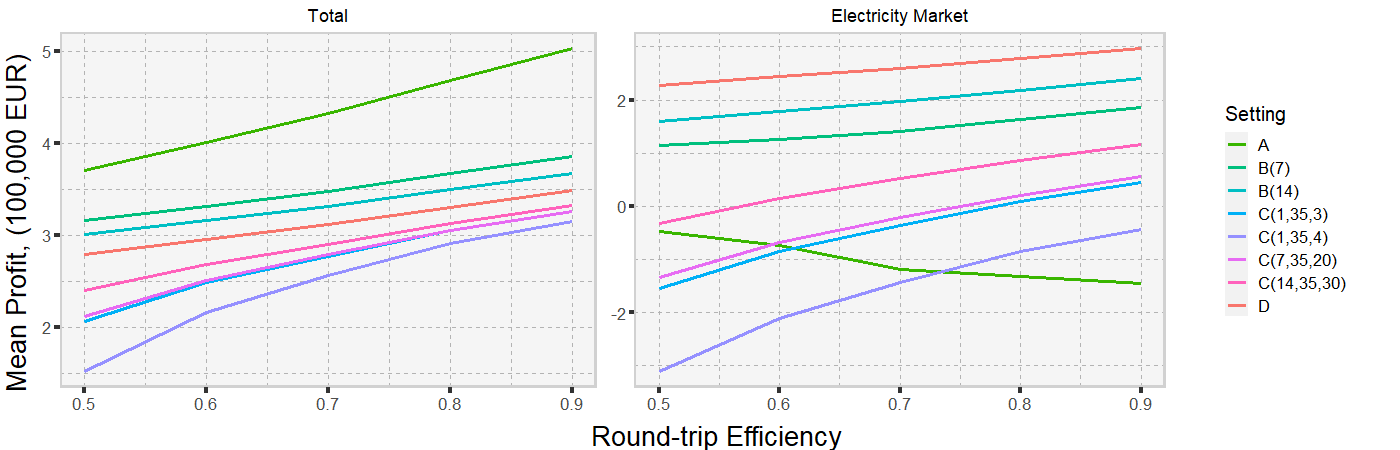}
    
    \caption{Impact of round-trip efficiency on the total expected rewards for the settings $A - D$.}
    \label{fig:efficiency}
\end{figure}

Research on technical developments in the electrolyzer technology is ongoing. In this set of experiments we examine the prospects of varying conversion efficiency rates. We vary the round trip efficiency~$\alpha$ between 0.5 and 0.9 in steps of 0.1. Recall that in our base system~$\alpha$ was set to 0.5. Figure~\ref{fig:efficiency} shows the expected total profit per year (top right) and its breakdown in profit due to selling hydrogen (top right), fulfilling PPA obligations (bottom left), and the profit for interacting with the market (bottom right). We included all relevant settings, i.e., we omit setting~$E$ in which no conversion takes place and we do not consider setting~$D(B)$ as setting~$D(H)$ with a conversion rate of 0.9 is equivalent.

Figure~\ref{fig:efficiency} shows how the expected yearly profit behaves for all settings that use HES if energy conversion can be done more efficiently. We show the increase in total profit and that related to trading on the power market. The increase in profit is largest for the fixed ($C$) settings. Under these settings, a fixed amount of hydrogen has to be sold. With low round trip efficiencies, it takes relatively more energy to generate the hydrogen. Therefore, relatively high amounts of energy have to be bought from the electricity market, leading to negative expected profits for low round trip efficiencies (right panel). When the round trip efficiency increases, less energy has to be bought, which increases both the profit made by trading at the electricity market (bottom left) and the total profit (top left). Similarly, the profits made by trading at the power market relatively increases under settings~$B$ when the round trip efficiency increases. Interestingly, this does not hold under setting~$A$. Under this setting, the increase in round trip efficiency makes the conversion to hydrogen less costly. Because complete flexibility is provided in terms of when to sell hydrogen, higher efficiency simply results in more hydrogen being sold rather than converting it back to electricity and engaging in the electricity market.

\subsection{Summary of Discussions}

From the presented results several points stand out. Green hydrogen plants that jointly produce renewable energy and operate hydrogen energy storage systems are operationally cost-efficient if sufficient flexibility to sell hydrogen can be provided. If a future hydrogen market will develop that behaves similar to the electricity market a substantial profit increase of~51\% can be obtained by coupling renewable energy production with an electrolyzer and a storage facility. On top of that, such an integrated system even increases profits by approximately~8\% by taking advantage of price fluctuations on both the electricity and hydrogen market, compared to a system with no hydrogen market opportunities but where energy conversions are almost without any losses -- a situation that is not realistic in the short-term. Nevertheless, this insight is crucial when assessing different technologies to invest on such as battery storage systems. However, we also note that although battery systems currently offer higher efficiency rates they are much less scalable than hydrogen storage.

Results also underline that hydrogen offtake agreements can be the motor of a transition towards large-scale green hydrogen production. Under prices equalling those of the power purchase agreements, it is shown that profitability decreases slightly (compared to a mature hydrogen market) but total hydrogen production is increased. If the latter is the goal, hydrogen offtake agreements will help to make steps in the energy transition. Moreover, if hydrogen prices are set to 11-13 \euro / MWh higher than the expected market price for electricity, profitability of operating a GHP under hydrogen offtake agreements can compensate for the energy losses during conversion. In the current, not-mature hydrogen market, market prices are exceeding these 11-13 \euro / MWh margins easily. 

Another observation relates to the importance of negotiating for the right PPA agreements. We see higher profits when selling to the PPA is spread out during periods rather than bulk selling at a specified due date. Likewise, the target delivery amount of the PPA should be carefully decided as both too high or too low amounts are detrimental to the overall profitability. 

Furthermore, results related to the inventory level throughout the year indicate that plant owners may benefit from seasonal storage agreements and thereby further reduce their storage costs.


\section{Conclusions}
\label{sec:conc}

In this paper, we study optimal control policies for renewable energy systems with co-located hydrogen storage with both electricity market interaction and the potential to sell hydrogen as a gas. This system, what we refer to as \textit{Green Hydrogen Plant}, can thus use its hydrogen storage to anticipate on fluctuating electricity prices (i.e., store now and sell later, or even buy from the market in case prices are low) and it can directly sell hydrogen as gas. We consider practical elements such as power purchase agreements (that dictate contractually binding electricity offtake at fixed time intervals) and hydrogen offtake agreements.

To study the profit-maximizing behaviour of the owner of a Green Hydrogen Factory, we modeled the stochastic sequential optimization problem on how much electricity to store (as hydrogen) buy, or sell, as well as the amount of hydrogen to sell, as a Markov decision process. We solved this system using backward dynamic programming to optimality. We are, thereby, the first to describe the optimal profit-maximizing control strategy of Green Hydrogen Plants.

In light of recent developments observed in the energy sector, our results show how a renewable energy producer can benefit from a storage facility and hydrogen market interactions. First, we show that incorporating hydrogen energy storage when hydrogen can be sold as a gas increases profits by~20\%. This is under the assumption that the generating hydrogen price process is similar to the electricity price. Future hydrogen prices may take different paths depending on many factors such as regulations and market demand. One expected outcome is for the hydrogen prices to become higher than the electricity prices with price drops in the power market with increased renewable energy penetration. This would lead to more favorable operating conditions for a Green Hydrogen Plant. Second, if the hydrogen distribution policies are fixed according to hydrogen offtake agreements, we observe that for a hydrogen price only marginally higher than the expected electricity price (in the range of \euro 11-13 per MWh) the losses due to low conversion rates can be compensated. Third, we show how plant operators can benefit from dynamic policies when setting their PPA agreements. Finally, our results highlight the benefit of coupled storage facilities providing additional profit opportunities up to~51\%. We also provide insights on to how to allocate these storage units throughout a year.

The opportunities for further research are numerous. A first research avenue could be to extend the market study to exploit interactions with neighboring regions considering import and export opportunities, which are expected to increase the gains of the considered systems. However, one should note the complexity of operating such units with inter-country regulations and rules. A second research avenue could be focusing on the design of a GHP in terms of its geographical location and its further integration in the power network. Summarizing, this paper provides a starting point for new research directions on the optimal control of future green hydrogen plants and can help in paving the way towards a carbon-neutral society.

\section*{Acknowledgements}

This project has received funding from the Fuel Cells and Hydrogen 2 Joint Undertaking under grant agreement No 875090, HEAVENN - Hydrogen Energy Applications in Valley Environments for Northern Netherlands. This Joint Undertaking receives support from the European Union’s Horizon 2020 research and innovation programme and Hydrogen Europe and Hydrogen Europe Research.

\appendix
\section{Wind Distributions}

In our experiments, we fitted daily wind speeds for each month of the year independently, similar as in \cite{schrotenboer2020mixed}. The fitted parameters are given in Table \ref{tab:weibull}.

\begin{table}[h!]
    \centering
    \begin{tabular}{l|rr}
        \toprule 
        Month & shape & scale \\ \midrule
        January & 2.514   & 6.816 \\
        February & 2.483  & 6.643 \\
        March & 2.566     & 6.413 \\
        April & 3.027     & 5.641 \\
        May & 3.388       & 5.670 \\
        June & 3.107      & 5.330 \\
        July & 3.144      & 5.142 \\
        August & 3.097    & 4.939 \\
        September & 2.641 & 5.223 \\
        October& 2.702    & 5.812 \\
        November & 2.695  & 5.959 \\
        December & 2.547  & 6.453 \\ \bottomrule
    \end{tabular}
    
    \caption{Parameters of the fitted wind-speed Weibull distributions}
    \label{tab:weibull}
\end{table}

\bibliographystyle{elsarticle-harv}
\bibliography{references}

\end{document}